\documentclass[prl,twocolumn,superscriptaddress,showpacs,floatfix]{revtex4}
\usepackage{graphicx}
\usepackage{wasysym}
\usepackage{bm}
\usepackage{epstopdf}

\begin{document}
\title{ Universal Thermoelectric Effect of Dirac Fermions in Graphene}
\author{Lijun Zhu}
\affiliation{Department of Physics and Astronomy, California State University,
Northridge, California 91330, USA}
\affiliation{Theoretical Division and Center for Nonlinear Studies, Los Alamos National Laboratory, 
Los Alamos, New Mexico 87545, USA}
\author{Rong Ma}	
\affiliation{Department of Physics and Astronomy, California State University, 
Northridge, California 91330, USA}
\affiliation{Department of Physics, Southeast University, Nanjing 210096, 
China}
\author{Li Sheng}
\affiliation{National Laboratory of Solid State Microstructures and 
Department of Physics, Nanjing University, Nanjing 210093, China }
\author{Mei Liu}
\affiliation{Department of Physics, Southeast University, Nanjing 210096, 
China}
\author{Dong-Ning Sheng}
\affiliation{Department of Physics and Astronomy, California State University, 
Northridge, California 91330, USA}

\date{\today}

\begin{abstract}
We numerically study the thermoelectric transports of Dirac fermions in 
graphene in the presence of a strong magnetic field and disorder. 
We find that the thermoelectric transport coefficients demonstrate universal 
behavior depending on the ratio between the temperature and the width of the 
disorder-broadened Landau levels(LLs). The  transverse thermoelectric 
conductivity $\alpha_{xy}$ reaches a universal quantum value at the center of 
each LL in the high temperature regime, and it has a linear temperature dependence 
at low temperatures. The calculated Nernst signal has a peak at the
central LL  with heights of the order of $k_B/e$, and changes sign near other 
LLs, while the thermopower has an opposite behavior, in good agreement with 
experimental data. The validity of the generalized Mott relation between the 
thermoelectric and electrical transport coefficients is verified in a wide 
range of temperatures.  
\end{abstract}

\maketitle

Graphene has attracted enormous interest due to its unique electronic 
properties associated with the two-dimensional (2D) Dirac-fermion excitations
~\cite{CastroNeto09}, including its thermoelectric properties~\cite{Kim09,
Shi09, Ong08}. The thermopower and Nernst coefficient, measuring the magnitude
of the longitudinal and transverse electric fields generated in response to an
applied temperature gradient, are very sensitive to the semimetal nature of graphene.
For instance, a large thermopower is expected near the Dirac point as $S_{xx} \sim 
T/E_F$ while $T$ and $E_F$ are the temperature and the Fermi energy, 
respectively. This has been observed in recent experiments
~\cite{Kim09,Shi09,Ong08}, where the maximum value of thermopower reaches
$90 \mu V/K$ at $T\approx 300K$.

In a magnetic field, the electronic states of graphene are quantized into 
Landau levels (LLs), as in the 2D semiconductor systems displaying the integer
quantum Hall effect (IQHE)~\cite{Prange87}. However, the Hall conductivity of
graphene obeys an unconventional quantization rule 
$\sigma_{xy}=4(n+1/2) e^2/h$, where $n$ is an 
integer~\cite{Novoselov05, Zhang05,Haldane88, Gusynin05}. 
For the conventional 2D IQHE systems, theories~\cite{Girvin82,Streda83,Jonson84,Oji84} 
predict that when the thermal activation dominates the broadening of LLs, all 
transport coefficients are universal functions of $E_F/\hbar \omega_c$ and 
$k_B T /\hbar \omega_c$, where $\hbar \omega_c$ is the LL quantization energy. 
$S_{xx}$ shows a series of peaks near the LL energies with height  
$ \ln2(k_B/e) /(n+1/2) $, which is independent of the magnetic field or
temperature. The Nernst signal $S_{xy}$ oscillates about zero near
the LLs and enhances as the strength of the impurity scattering
increases. The experimental results on graphene agree with these
asymptotic behaviors except at the central LL, where $S_{xy}$ has a
peak instead with maximum value about $40 \mu V/K$ while $S_{xx}$ 
becomes oscillatory~\cite{Kim09,Shi09,Ong08}. 
The unusual behavior of $S_{xy}$  and $S_{xx}$ near the central LL has not 
been understood.

While thermoelectric transports depend crucially on impurity scattering as 
well as thermal activation, the study of disorder effect on thermoelectric 
transports in graphene is still lacking. Another important question is to what 
extent the well-known Mott relation between thermoelectric and electrical 
transport coefficients [cf. Eq.(\ref{eq:Mott-relation})] is applicable for
this system. In this Letter, we carry out a numerical study to address all the 
above issues.  We show that  thermoelectric transport coefficients are 
universal functions of the ratio between the temperature and the 
disorder-induced LL width, and display different asymptotic behaviors in 
different temperature regions, in agreement with experimental data. 
Our study also reveals that the distinct behaviors in the central LL
is an intrisic property of the Dirac point where both particle
and hole LLs coexist. Furthermore,  the generalized Mott relation is shown 
to be valid for a wide range of temperatures.

We consider a rectangular sample of a 2D graphene sheet consisting of carbon 
atoms on a honeycomb lattice~\cite{Sheng05}. Besides the 
Anderson-type random disorder considered in Ref.\cite{Sheng05}, we also model
charged impurities in substrate, randomly located in a plane at a distance $d$,
either above or below the graphene sheet with a long-range Coulomb scattering 
potential. The latter type of disorder is known~\cite{Adam07} to give a more 
satisfactory interpretation of the transport properties of graphene in the 
absence of magnetic fields. When a magnetic field is applied perpendicular to
the graphene plane, the Hamiltonian can be written in the tight-binding form
\begin{equation}
H=-t\sum\limits_{\langle ij\rangle \sigma }e^{ia_{ij}}
c_{i\sigma }^{\dagger }c_{j\sigma}
+\sum_{i\sigma}w_ic_{i\sigma }^{\dagger }c_{i\sigma},
\label{eq:Hamiltonian}
\end{equation}
where $c_{i\sigma }^{+}$ ($c_{i\sigma }$) creates (annihilates) a $\pi$ 
electron of spin $\sigma $ on lattice site $i$. $t$ is the
nearest-neighbor hopping integral with an additional phase factor $a_{ij}$ due to the applied 
magnetic field $B$. 
The magnetic flux per hexagon 
$\phi =\sum_{{\small {\mbox{\hexagon}}}}a_{ij}=\frac{2\pi }{M}$ with $M$ an integer ~\cite{Sheng05}. 
For Anderson-type disorder, $w_i$ is randomly distributed between $[-W/2,W/2]$
with $W$ as the disorder strength. For charged impurities, $w_i=
-\frac{Ze^2}{\kappa}\sum_{\alpha}1/\sqrt{({\bf r}_i-{\bf R}_{\alpha})^2+d^2}$,
where $Ze$ is the charge carried by the impurities, $\kappa$ is the effective 
background lattice dielectric constant, and ${\bf r}_i$ and ${\bf R}_{\alpha}$
are the planar positions of site $i$ and impurity $\alpha$, respectively. All 
the properties of the substrate (or vacuum in the case of suspended graphene) 
can be absorbed into a dimensionless parameter $r_s= Ze^2/(\kappa \hbar v_F)$, 
where $v_F$ is the Fermi velocity of the electrons.
For simplicity, in the following calculation, we fix the value of distance 
$d=1$,  randomly distribute impurities with the density as $1\%$ of the total sites, 
and tune $r_s$ to control the impurity scattering strength.

In the linear response regime, the charge current in response to an electric 
field or a temperature gradient can be written as  
${\bf J} = {\hat \sigma} {\bf E} + {\hat \alpha} (-\nabla T)$, 
where ${\hat \sigma}$ and ${\hat \alpha}$ are the electrical and thermoelectric
conductivity tensors, respectively.  These transport coefficients  can be
calculated with Kubo formula once we obtain all the eigenstates of the 
Hamiltonian (in our calculation, $\sigma _{xx}$ is obtained based on the 
calculation of the Thouless number \cite{Ma08}). In practice, we can first 
calculate the $T=0$ conductivities $\sigma_{ji}(E_F)$, and then use the 
relation~\cite{Jonson84}
\begin{eqnarray}
\sigma_{ji}(E_F, T) &=& \int d\epsilon \,\sigma_{ji}(\epsilon) 
\left ( - {\partial f(\epsilon) \over \partial \epsilon } \right), \nonumber \\
\alpha_{ji}(E_F, T) &=& {-1\over eT} \int d\epsilon\, \sigma_{ji}(\epsilon) 
(\epsilon-E_F) \left ( - {\partial f(\epsilon) \over \partial \epsilon } \right),
\label{eq:conductance-finiteT}
\end{eqnarray}
to obtain the finite temperature electrical and thermoelectric conductivity 
tensors. Here $f(x) = 1/[e^{(x-E_F)/k_B T}+1]$ is the Fermi distribution 
function.
At low temperatures, the second equation can be approximated as
\begin{equation}
\alpha_{ji}(E_F, T) =-\frac {\pi^2k_B^2T}{3e}\left. 
\frac {d\sigma_{ji}(\epsilon, T)}{d\epsilon} \right|_{\epsilon =E_F},
\label{eq:Mott-relation}
\end{equation}
which is the generalized Mott relation~\cite{Jonson84,Oji84}. 

\begin{figure}[tbp]
\par
\includegraphics[width=\columnwidth]{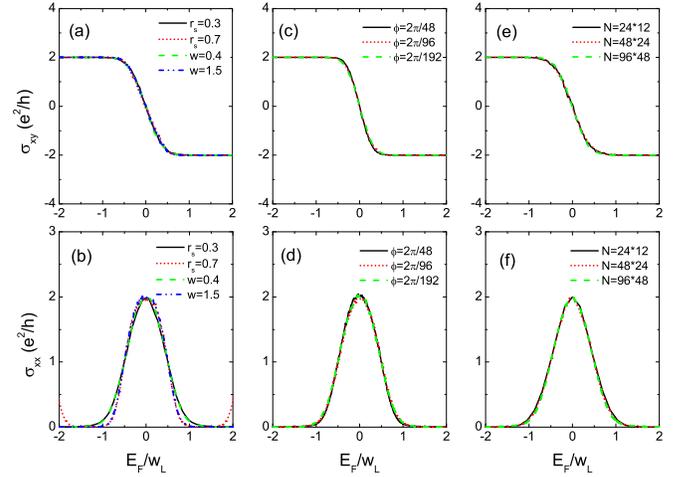}
\caption{ (color online). Zero temperature $\sigma _{xy}$ and  $\sigma_{xx}$ as 
functions of the normalized Fermi energy near the Dirac point with $W_{L}$ as 
the width of the central LL. (a) and (b) compare results for short-range random 
potential (with various $W$) and long-range Coulomb potential (with various 
$r_s$), where the system size $N=96\times 48$ and the magnetic flux 
$\protect\phi=\frac{2\protect\pi}{48}$. (c) and (d) compare results for
three different strengths of magnetic flux, with the same system size 
$N=96\times 48$ and the same impurity configuration, where a long-range 
scattering potential with $r_s=0.4$ is assumed. (e) and (f) are for three different 
system sizes with $r_s=0.3$ and $\phi =\frac{2\pi}{24}$.}
\label{fig:conductance}
\end{figure}

In Fig.\ref{fig:conductance}, we show the calculated Hall conductivity
$\sigma _{xy} $ and longitudinal conductivity $\sigma _{xx}$  at $T=0$ as 
functions of the Fermi energy near the Dirac point. From Fig. 
\ref{fig:conductance} (a) and (b), we observe that the results for the two different 
types of disorder are very similar. The Hall 
conductivity exhibits two well-quantized plateaus $-2e^2/h$ and $2e^2/h$, being 
consistent with the quantization rule $\sigma_{xy}=4(n+1/2)e^2/h$. The direct 
transition between the two plateaus is accompanied by a pronounced peak in
the longitudinal conductivity $\sigma_{xx}$ with the maximum value
$2e^2/h$~\cite{footnote2}. Remarkably, once we scale the energy with the width 
of the central LL ($W_L$), which is determined by the full-width at half-maximum 
of the $\sigma_{xx}$ peak, all results fall into a single curve, though there are 
small deviations for $\sigma_{xx}$ at the peak tails. This is rather in 
accordance with the scaling theory on the  quantum Hall liquid  to insulator 
transition~\cite{Sondhi97}, as $W_L$ corresponds to an energy scale where the 
electron localization length (correlation length) is comparable to the system 
size. The scaling is further tested for different magnetic fields and different 
system sizes, as shown in Figs.~\ref{fig:conductance}(c-d) and 
Figs.~\ref{fig:conductance}(e-f), respectively.  We conclude that, when 
$E_F,W_L \ll \hbar \omega_c$, $\sigma_{xx}$ and $\sigma_{xy}$ are universal 
functions of a single parameter $E_F/W_L$ .
It is noteworthy that the universal curves shown in Fig. 1 are for disorder strengths 
smaller than the critical values, which is relevant to the 
experiments~\cite{Kim09,Shi09,Ong08}. With stronger disorder strength, an 
insulating state may appear at the Dirac point~\cite{Sheng05}. 

\begin{figure}[tbh]
\includegraphics[width=\columnwidth]{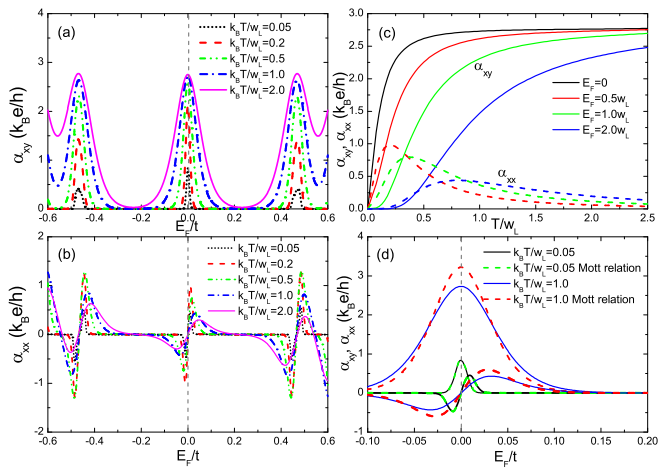}
\caption{(color online). Thermoelectric conductivities at finite temperatures. 
(a) and (b) show $\alpha_{xy}(E_F, T)$ and $\alpha_{xx}(E_F,T)$ as functions of 
the Fermi energy at different temperatures. (c) shows the temperature dependence 
of $\alpha_{xy}(E_F,T)$ and $\alpha_{xx}(E_F, T)$  for certain fixed Fermi 
energies.  (d) compares the results from numerical calculations and from the 
generalized Mott relation at two characteristic temperatures, $k_{B}T/W_L=0.05$ 
and $k_BT/W_L=1$. The other parameters are taken to be $N=96\times48$, 
$\phi=2\pi/48$, and $r_s =0.3$ with $W_L/t = 0.0195$. }
\label{fig:thermoelectric-finiteT}
\end{figure}

In Fig. \ref{fig:thermoelectric-finiteT}, we show the results of thermoelectric 
conductivities $\alpha_{xy}$ and $\alpha_{xx}$ at finite temperatures.
As seen from Fig.\ref{fig:thermoelectric-finiteT}(a) and (b),
$\alpha_{xy}$ displays a series of peaks near the LL energies, while 
${\alpha_{xx}}$ oscillates and changes sign at the LL energies. These behaviors
are similar to those in the conventional IQHE systems~\cite{Jonson84}, but some
important differences exist. First, at low temperatures, the peak of 
$\alpha_{xy}$ at the central LL is higher and narrower than at the other LLs, 
which indicates that the impurity scattering has a different effect on the 
central LL and the rest LLs. Second, $\alpha_{xy}$ and $\alpha_{xx}$ are 
symmetric and antisymmetric about the Dirac point (zero energy), respectively,
as they are even and odd functions of the Fermi energy, rather than periodic 
functions.  We also find that, depending on the relative strength between 
$k_{B}T$ and $W_L$, the thermoelectric conductivities show different universal 
behaviors. When $k_{B}T \ll W_L$ and $E_F \ll W_L$, we find that both $\alpha_{xy}$ 
and  $\alpha_{xx}$ are linear in $T$  (verified by a log-log  plot to the lowest  
temperature accessible by numerics). 
This indicates that within the mobility edge where extended states dominate, 
the diffusive transports play important roles as in semiclassical Drude-Zener regime. 
When the Fermi energy falls deep inside the mobility gap 
or $k_{B}T$ becomes comparable to or greater 
than $W_L$,   thermal activation dominates. In 
this regime, $\alpha_{xx}$ assumes the Arrhenius form 
$(1/T)e^{-E_F/k_{B}T}$.  
Meanwhile, the heights of the peaks in $\alpha_{xy}$ for all LLs saturate to a 
value $2.77 k_B e/h$, as seen from Fig.\ref{fig:thermoelectric-finiteT}(c). This 
value matches exactly the universal number $(\ln 2) k_B e/h$ predicted for the 
conventional IQHE systems in the case where thermal activation 
dominates~\cite{Jonson84, Oji84}, with an additional degeneracy factor $4$.

To examine the validity of the generalized Mott relation, we compare the above
results with those calculated from  Eq.(\ref{eq:Mott-relation}), as shown in 
Fig.\ref{fig:thermoelectric-finiteT}(d). The Mott relation, which was 
historically derived from the semiclassical Boltzmann equation, is a 
low-temperature approximation and predicts that  thermoelectric conductivities 
are linear in temperature. This is in agreement with our low-temperature results. 
At high temperatures,  thermoelectric conductivities deviate from the 
linear-$T$ dependence. However, if we take into account the finite temperature 
values of  electrical conductivities, the Mott relation still predicts the 
correct asymptotic behavior. This is due to the fact that the conductivities 
display universal dependence on $E_F/k_{B}T$, either a power law or an 
exponential function in different regimes, which can all be captured by the Mott 
relation. This proves that the generalized Mott relation is asymptotically valid 
in Landau-quantized systems, as suggested in Ref.~\cite{Jonson84}.

\begin{figure}[tbh]
\includegraphics[width=0.75\columnwidth]{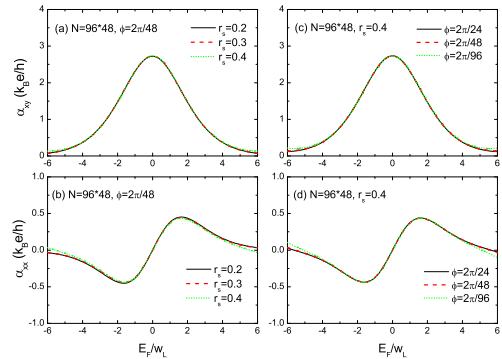}
\caption{ (color online). Universal behaviors of $\alpha_{xx}$ and $\alpha_{xy}$ 
near the central LL at $k_{B}T/ W_L=1$.  (a)(b) show them as functions of the 
renormalized Fermi energy for three disorder strengths, and (c)(d) compare results for 
different magnetic fields. The parameters chosen are shown in the figure. }
\label{fig:alpha-universal}
\end{figure}

Given the single-parameter scaling 
behaviors of the $T=0$ conductivities,  it is straightforward to  show with
Eq.(\ref{eq:conductance-finiteT}) that $\sigma_{ji}$ and $\alpha_{ji}$ at finite 
temperatures are universal functions of $W_L/k_BT$ and $E_F/k_B T$, or 
\begin{equation}
{\cal L} (E_F, T) =  {\cal L}_0 S_{\cal L}\left({ E_F/ W_L}, 
{W_L / k_B T} \right),
\label{eq:universal}
\end{equation}
where ${\cal L}$ stands for one of the transport coefficients and 
$S_{\cal L}(x, y)$ is a universal function. We can directly verify the universal 
relations Eq.(\ref{eq:universal}). Here, we pick a temperature $k_{B}T = W_L$, 
corresponding to $S_{\cal L}(x, 1)$, and compare the results around the central LL for 
different disorder strengths  and different magnetic fields, as shown in Fig. 
\ref{fig:alpha-universal}.  Indeed, once we scale the Fermi energy with $W_L$,  
$\alpha_{xx}$ and $\alpha_{xy}$ as functions of $E_F/W_L$ collapse into the same 
corresponding curves.

\begin{figure}[tbh]
\includegraphics[width=0.75\columnwidth]{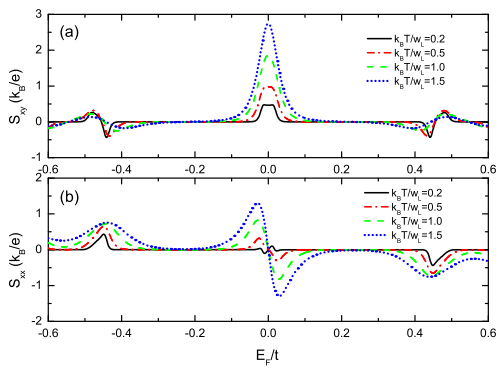}
\caption{ (color online). Calculated (a) $S_{xx}$ and (b) $S_{xy}$ as functions 
of the Fermi energy for different temperatures. The parameters chosen here are 
the same as in Fig.\ref{fig:thermoelectric-finiteT}. }
\label{fig:tep-nernst}
\end{figure}

We note that the asymptotic behaviors and the values of the peak heights
of our calculated $\alpha_{xx}$ and $\alpha_{xy}$ are in agreement with the 
experimental results~\cite{Ong08}. We further calculate the thermopower $S_{xx}$
and the Nernst signal $S_{xy}$ using~\cite{Jonson84}
\begin{equation}
S_{ij} = E_j/\nabla_i T= \sum_{k=x,y} [{\hat \sigma}^{-1}]_{ik} \alpha_{kj},
\label{eq:thermoelectric}
\end{equation}
which are directly determined in experiments by measuring the responsive electric fields. 
The results for three central LLs are shown in Fig. {\ref{fig:tep-nernst}. 
$S_{xy}$ ($S_{xx}$) has a peak at the central LL (the other LLs), and changes 
sign near the other LLs (the central LL). The height of the $S_{xx}$ peak at 
$n=-1$ LL is found to be $37\mu V/K$ for $k_BT=0.2W_L$ and $56\mu V/K$ for 
$k_BT=0.5W_L$, which is in agreement with the maximum measured value 
$48\mu V/K$\cite{Ong08}. This is also in agreement with the theory predication that, 
in the absence of disorder and at low temperatures, the 
peak value of $S_{xx}$ is dominated by $\alpha_{xy}/\sigma_{xy}$ and takes a 
universal value $(2/3)\ln2 k_B/e \approx 40 \mu V/K $.  In the presence of 
disorder and at finite temperatures, the peak value is slightly bigger and the 
peak position is shifted toward $E_F=0$. At $E_F=0$, both $\sigma_{xy}$ and 
$\alpha_{xx}$ vanish, leading to a vanishing $S_{xx}$.  Around the zero energy, 
because $\sigma_{xx} \alpha_{xx}$ and $\sigma_{xy}\alpha_{xy}$ have opposite 
signs, depending on their relative magnitudes, $S_{xx}$ increases or decreases
when the Fermi energy is increased passing the Dirac point.  In our calculations, we find 
that the second term is always dominant, which is different from the experimental 
observation. This might be related to the unexpected large value of  
$\sigma_{xx}\sim 6 e^2/h$ observed in experiments.  Figure \ {\ref{fig:tep-nernst} 
actually shows the result when we rescale $\sigma_{xx}$ to the experimental value 
instead of the theoretical result $2e^2/h$. We can then obtain the same 
asymptotic behavior as in experiments for $k_{B}T=0.2W_{L}$. On the other hand, 
$S_{xy}$ has a peak structure, which is dominated by 
$\alpha_{xy}/\sigma_{xx}$. With the rescaled $\sigma_{xx}$, we find that the peak 
height is $40 \mu V/K$ at $k_BT=0.2W_L$, which is comparable with the 
experimental value $20-40 \mu V/K$. 
The distinct thermoelectric behaviors near the central LL can be traced down to 
the Berry phase anomaly of the Dirac point. 
The central LL in graphene in fact consists of two degenerate LLs,  
one from particles and another from holes, which are protected by the particle-hole symmetry. 
While this makes no distinction in electrical transports when the energy is tuned 
through the LL,  electrons and holes contribute differently in thermoelectric transports
than in electrical transports.  Here, the contributions of equal numbers of 
electrons and holes moving along the same thermal gradient direction cancel with 
each other  in $S_{xx}$  and are additive in $S_{xy}$.  Other LLs, without this property, 
have similar behaviors as in conventional IQHE systems.

In summary, we have investigated the thermoelectric transports in graphene
by a numerical study on the lattice model in the presence of both disorder
and a magnetic field and obtain results in agreement with experiments.  

This work is supported by the U.S. DOE Grant No. DE-FG02-06ER46305 (L.Z, D.N.S), 
the U.S. DOE at LANL under Contract No. DE-AC52-06NA25396(L.Z), 
the NSF Grant Nos. DMR-0605696 and 0906816 (R.M, D.N.S), the NSFC Grant No. 10874066, 
the National Basic Research Program of China under Grant
Nos. 2007CB925104 and 2009CB929504 (L.S), 
and the doctoral foundation of Chinese Universities under
Grant No. 20060286044 (M.L).
We also acknowledge partial support from  Princeton MRSEC Grant No.DMR-0819860,
the KITP through the NSF Grant No. PHY05-51164, 
the State Scholarship Fund from the China Scholarship Council and
the Scientific Research Foundation of Graduate School of Southeast 
University of China (R.M).

\end{document}